# "Study of an interaction between the jet and an interstellar medium of M87 with a spectral analysis by using *Chandra*"

S.Osone
Chiba, Japan
e-mail : osonesatoko@gmail.com

Abstract
Both a thermal component and a non thermal component as an interaction between the jet and an interstellar medium are studied with an exposure time of 800 ks of *Chandra*. It is confirmed that an X ray energy spectra for the nucleus and the knot D is well described with a power law model as synchrotron emission. It is found that a power law model is rejected statistically for HST-1 and the knot A and an X ray energy spectra is well described with a combination model of a power law and a thermal bremsstrahlung of temperature of 0.2 keV and metal abundance of 0.00. There is a possibility that gas in a jet is heated by a shock as interaction between the jet and an interstellar medium and it is observed as thermal emission. The flux of a non thermal bremsstrahlung from the jet as an interaction between accelerated electrons and an interstellar medium is calculated from an X ray result and it is found that there is no contribution from both nucleus and the knot HST-1 to the observed flux with *Fermi*.

## 1 Introduction

M87 is a close radio galaxy and is a center of the Virgo cluster. The distance is 16.8 Mpc (Blakeslee et al. 2009 ). M87 is famous for a blackhole shadow observation by EHT(EHT 2019).

M87 is observed over a wide spectrum from a radio frequency to TeV gamma ray. M87 has an inclined jet of 20 arc second length. The angular resolution is micro second in radio, 0.7 arc second in optical band and 0.5 arc second in X ray by *Chandra*, 5.2 arc minutes in GeV gamma ray by *Fermi* and 2 arc minutes in TeV gamma ray by air Cherenkov detector as CTA, HESS, LHASSO, MAGIC and VERITAS, 0.1 degree in TeV gamma ray by water Cherenkov detector as HWAC and LHASSO, and 0.5 degree in TeV gamma ray by air shower arrays as LHASSO and Tibet AS $\gamma$ . The jet cannot be resolved in a high-energy gamma ray.

Therefore, the origin of TeV gamma ray is studied with both a timescale of a flux variability and a correlation of flux with a simultaneous observation. A flux variability of a timescale $t_{val}$ of 2 d in TeV gamma ray was observed (Aharonian et al. 2006). The size of the emitting area is given by $ct_{val}\delta = 3.1\text{x}10^{16} (\delta / 6)$ cm (0.01 pc). Here, $c$ is a speed of a light and $\delta$ is a beaming factor. Therefore, they conclude that the origin of TeV gamma ray is the nucleus. A correlation between the nucleus with a radio of 43 GHz and TeV gamma ray with VERITAS also show that the origin of TeV gamma ray is nucleus (Acciari et al. 2009). A correlation between X-rays of the nucleus and TeV gamma ray was detected in 2008 and 2010, while that between X-ray of the knot HST-1 and TeV gamma ray was detected in 2005 (Abramowaski et al. 2012). The knot HST-1 is also an origin of TeV gamma ray. Radio, optical, and X-ray observations of the nucleus are used for a study of multi wavelength energy spectra.

Non-simultaneous multi-wavelength energy spectra are described with various models (Reiger & Aharonian 2012). The models are classified into three categories: leptonic models, hadronic models and others. Leptonic models can be a homogeneous one-zone synchrotron self Compton model (Finke, Dermer and Bottcher 2008), a decelerating jet model (Georganopoulus, Perlman and Kazanas 2005), an electron of a jet model (Stawarz et al. 2005, 2006), a multi-blob model (Lenain et al. 2008), a spine-sheath layer model(Tavecchio & Ghisellini 2008), a jet in a jet model (Giannios, Uzdensky and Begelman 2010) and an external inverse Compton model (Cui et al. 2012). Hadronic models can be a proton synchrotron model(Reimer, Protheroe and Donea 2004), an interaction model between protons and an interstellar medium (Pfrommer & Enβlin 2003), and an interaction between protons and a cloud injected to the jet(Barkov, Ramon and Aharonian 2012). Others include a lepto-hadronic model (Reynoso, Medina and Romero 2011), a dark matter annihilation model(Baltz et al. 2000; Saxena et al. 2011) and a magnetospheric model (Levinson & Rieger 2011; Broderick & Tchekhovskoy 2015).

Recently, a detection of a diffused thermal component of 1 keV for a radio lobe of radio galaxies, Cen A (Stawarz et al. 2013; O'sullivan et al. 2013) and Fornax A(Seta, Tashiro and Inoue 2013) was reported.

There is a possibility of a thermal component for not only the radio lobe of a jet, but also a jet itself. For SS433, there is an interaction bwteen a jet and an interstellar medium.

Monocular cloud were found along an X-ray jet for SS433 (Yamamoto et al. 2022), the correlation between molecular cloud and an absorption in X-ray was found(Kayama et al. 2022) and thermal emission were detected from the jet (eg. Shidatsu et al. 2025).

Dainotti et al. (2012) studied an interaction between high energy cosmic rays and an interstellar medium about the M87 jet with a morphological analysis by *Chandra* and pointed out the decreasing soft X-ray emission in a surrounding from the knot E to the knot F of the M87 jet as a cosmic ray cocoon. There is a possibility that soft X-ray is absorbed in the compressed interstellar medium.

The X-ray energy spectra of the M87 jet has been analyzed with *Chandra* (Wilson & Yang 2002; Marshall et al. 2002; Perlman & Wilson 2005; Sun et al. 2018). They fitted energy spectra with a power law model and obtained an acceptable fit.

Hot gas of the Virgo cluster centered at M87 has been reported with *XMM* (Belsole et al. 2001). The background for the jet is the sum of Cosmic X-ray background, galactic emission, solar activity, Non X-ray background and hot gas of Virgo cluster. The emission of hot gas of Virgo cluster depends on a distance from the center. Therefore, background should be taken according to a distance from the nucleus. Perlman & Wilson (2005) and Sun et al.(2018) takes two neighboring regions, a south and a north of the whole jet as a common background for all part of the jet. This is not correct.

Both Wilson & Yang (2002) and Perlman & Wilson (2005) use only obsID 1808 with an exposure time of 13 ks for CCD. Marshall et al. (2002) use data with an exposure time of 38 ks for High Energy Transmission Grating Spectrometer. Its effective area is about one tenth of that of CCD and this statistics corresponds to 4 ks with CCD. There are two problems with poor statistics. Any models tends to be acceptable with poor statistics. The other is contamination of hot gas with poor statistics. There is some contamination of hot gas of the cluster shown as a line feature on the energy spectra which subtract background for obsID 1808. The subtraction of background is not enough with an exposure time of 13 ks. With more statistics, this contamination is expected to be less. This is the reason why high statistics by using all archive is needed.

In this thesis, statistics is improved largely with an exposure time of about 800 ks and a correct subtraction of background is done with *Chandra*. Statistics is almost same with that of Sun et al.(2018).

Both a thermal component and a non thermal component as an interaction between the jet and an interstellar medium of M87 with a spectral analysis using *Chandra* are studied. An additional thermal component to a synchrotron emission from the M87 jet as heated gas by a shock is examined. The flux of non thermal bremsstrahlung as an interaction between accelerated electrons and an interstellar medium is calculated from an X-ray result and is compared with the observed flux with *Fermi*.

## 2 Spectral analysis with *Chandra*

The detector used is CCD. The image is taken from 2 dimension arrays of CCD and an X-ray energy spectra is taken from a deposit energy in CCD. I use archival data from 2000 to 2018. The target is the nucleus, the knot HST-1, the knot D and the knot A. These targets are saturated in a frame time of 3.2 sec. Therefore, a frame time of 0.4 sec is used. Data observed from 2000 Jul. to 2014 Dec. is called as a former data set and data observed from 2015 Mar. to 2018 Apr. is called as a latter data set. CIAO software of 4.15 and caldb 4.10.2 are used. The region file which show a position of a source and its radius and that which show a position of background and its radius are made by ds9 tool. The radius of circle as source region and a background region is 0.5 arc second for nucleus, 0.6 arc second for HST-1, 0.75 arcsecond for knot D and 1 arc second for knot A.

### 2.1 Background

Background for the jet of M87 is cosmic X-ray Background, detector background, galactic emission, solar acitivity and hot gas of Virgo cluster. X-ray emission of hot gas of Virgo cluster depends on the distance from the nucleus of M87(Bohringer et al. 2001) and there is a soft X-ray dip outside between knot E and knot F(Dainotti et al. 2012).

Therefore, background is not common for each part of the jet. Backrgound region is taken as neighbour to source region, north and south with same radius and with almost same distance from the nucleus for nucleus, HST1, knot D and knot A.

For knot A, another background region is also made which only south region is taken as background because the north region for knot A may contain the soft X-ray dip area.

A position of a source region for obsID 18232 is shown in table1. The image of obsID18232 made by ds9 tool is shown in figure1. The exposure time of obsID 18232 is 18 ks.

A position of each knot is changed from data to data by apparent velocity and precession. A position of an extracted region is decided by image for each data set. A radius of an extracted region is same by a data set.

### 2.2 pile up treatment

Nucleus and HST-1 are sometimes piled up heavily in a former data set. Heavy pile up distorts X-ray energy spectra. I select data with following three steps. At step1, I check an image by ds9 tool. The apparent pile up line is sometimes originated from nucleus or HST-1. There data is not used for analysis of nucleus and HST-1. The pile up line is sometimes overlapped with a source region or a background region for knot D and knot A. these data set is not used for analysis of knot D and knot A. At step2, I select data which nucleus and HST-1 are discriminated apparently. At step3, I check a pile up in count rate by a pileup_map command. This command makes a pile up information from an event data and the count/frame is calculated for a given region file by a dmstat command. I show count/frame for each knot in figure 2. The count/frame = 0.1, 0.2 corresponds to 5%, 10% pile up fraction, respectively. I select data with 5% pile up fraction. However, total observation time with a pile up fraction 5% is small for nucleus and HST-1 in a former data set. Therefore, data selected with a 10% pile up contamination are also used for nucleus and HST-1. For a latter data set, pile up fraction is less than a 5% for all data of all knots. The used observation log is shown in table 2 for nucleus, HST-1, knot D and knot A, respectively.

2.3 energy spectra

An energy spectra of source region and background region are made by a specextract command from fits data and a region file, respectively. The effective area (arf) and energy response(rmf) are also made by specextract command from fits data and a region file. The arf is made with no weight of a count rate and a correction by PSF which are suitable conditions for a point source analysis. Arf and rmf are made for only source region. Arf and rmf are multiplied as rsp for each data set and rsp for summed data set is weighted with an exposure time. An area of background region is two times as large as that of source region. When an energy spectra is made, a keyword BACKSCAL in a spectra file for background is set two time as large as that in a spectra file for source. Data points in energy spectra of a source are binned so that minimun counts per a bin are above 15.

The summed energy spectra which subtract background is described with no line feature and a contamination of hot gas of cluster is excluded successfully.

2.4 model fitting

CCD has a sensitivity from 0.2 keV to 10 keV. There is a quantum efficiency degradation by a contamination of an optical filter. An energy range above 0.3 keV, especially above 0.5 keV is well calibrated. Therefore, a lower limit of an energy in energy spectra is set as 0.5 keV.

XSPEC tool is used for an energy spectra analysis. At first, a power law model as synchrotron emission of accelerated electrons is used. Here, an absorption in soft X-ray is caused by a photo electric effect of a neutral material in a line of a sight from us to M87. The column density by a 21 cm radio observation is 1.27x10$^{-20}$cm$^{-2}$ (Bekhti et al 2016). An absorption is applied to a power law. A phabs model is used as an absorption. At first, a column density is set free. When a column density is lower than the 21 cm value, a column density is fixed to the 21 cm value and energy spectra is fitted with a power law. When a power law model is not reasonable, an APEC model as thermal bremsstrahlung with a metal is added to a power law model and energy spectra is fitted. Here, an absorption is also applied to an APEC model.

## 3 Result

### 3.1 Former data set

Fitting results are shown in table 3 and 4. For nucleus and HST-1, knot D, an energy spectra is well described with a power law model. For knot A, an energy spectra cannot be described with a power law model. $\chi^2$=462.19 for d.o.f =295 for a power law model show a chance probability of 1.6x$10^{-9}$. There is soft excess. Therefore, when an APEC model is added to a power law model, a reasonable fitting is obtained. The temperature of 0.2 keV and a metal abundance of 0.00 are obtained. In an APEC model, a relative abundance is fixed to a value of Grevesse & Anders (1989). A relative abundance is cannot be free because of large uncertainty. For knot A, when background is taken for only south, $\chi^2$=443.99 for d.o.f=295 for a power law model show the chance probability of 4.3x$10^{-8}$ When energy spectra is fitted with a combination model of a power law and APEC, a temperature of 0.2 keV and a metal abundance of 0.00 are obtained.

### 3.2 Latter data set

Fitting results are shown in table 5 and 6. For All knots, energy spectra are well described with a power law model. For knot A, $\chi^2$= 352.49 for d.o.f=299 for a power law model show a chance probability of 1.8x$10^{-2}$. For knot A, when background is taken for only south, $\chi^2$=354.96 for d.o.f=299 for a power law model show a chance probability of 1.4x$10^{-2}$.

### 3.3 All data set

Fitting results are shown in table 7 and 8. For nucleus and knot D, energy spectra is well described with a power law model. For HST-1 with 10% pile up fraction, an energy spectra cannot be described with a power law model. $\chi^2$=378.86 for d.o.f= 288 for a power law model show the chance probability of 2.6x$10^{-4}$. There is a soft excess. When an APEC model is added to a power law model, a reasonable result is obtained. The temperature of 0.2 keV and a metal abundance of 0.00 are obtained. In an APEC model, a relative abundance is fixed to a value of Grevesse & Anders (1989). A relative abundance is cannot be free because of large uncertainty. For knot A, an energy spectra cannot be described with a power law model. There is a soft excess as shown in figure 2 (top). $\chi^2$=628.76 for d.o.f=355 for a power law model show the chance probability of $1.6\text{x}10^{-17}$. A power law model is rejected statistically. When an APEC model is added to a power law model, a reasonable result is obtained as shown in figure 2 (bottom). The temperature of 0.2 keV and a metal abundance of 0.00 are obtained. The flux ratio of an APEC model to total is 22%. In an APEC model, a relative abundance is fixed to a value of Grevesse & Anders (1989). A relative abundance is cannot be free because of large uncertainty. For knot A, when background is taken for only south, $\chi^2$=645.54 for d.o.f=355 for a power law model show a chance probability of $3.7\text{x}10^{-19}$. When energy spectra is fitted with a combination model of a power law and an APEC, a temperature of 0.2 keV and a metal abundance of 0.00 are obtained. The metal abundance is low compared with hot gas of Virgo cluster which metal abundance is 1 solar (Belsole et al 2001). When a metal abundance is fixed to 0.1 solar or 1 solar, these models cannot be rejected as shown in table 9.

### 3.4 Comparison with *XMM* result

Bohringer et al. (2001) analyzed the X-ray energy spectra of an outer part of the jet with *XMM*. This location is 11 arc second apart from the nucleus, which is probably the knot A. The energy spectra of the jet is well described with an absorbed power law for both PN CCD and MOS CCD of *XMM*. Here, a column density is fixed to a 21 cm radio observation value.

There are two problems in their analysis. One is background and the other is signal to noise ratio. The location of a background is 20.8 arc second in north and 17.7 arc second in south for PN CCD, 22 arc second for MOS CCD from the nucleus. The distance is far from the source region. This is not correct. The exposure time is about 30 ks for both PN CCD and MOS CCD of *XMM* as compared with about 800 ks of *Chandra*. The effective area of PN CCD and MOS CCD of *XMM* are 1400 $cm^2$ and 800 $cm^2$ as compared with 300 $cm^2$ of CCD of *Chandra*. A FWHM of angular resolution of *XMM* is poor as 4 arc second and the radius of an extracted region is 4 arc second. The radius of a bright spot as the knot A is 1 arc second with *Chandra*. Therefore, the signal to noise ratio of PN CCD and MOS CCD of *XMM* is 1/10 and 1/14 of that of *Chandra* respectively. It is not strange that thermal component is not needed for an observation result with *XMM*.

## 4 Discussion

### 4.1 Physical values

For a former data set, a latter data set and all data, three physical values are calculated respectively: a neutral density $N$ from a column density $N_H$, a plasma density $n_i n_e$ from a normalization of an APEC model and a density of accelerated electrons $K'$ from both 1 keV flux and photon index of a power law. Three physical values are shown in table 10.

### 4.1.1 Neutral density

A column density by a 21 cm radio observation is due to our galaxy. The difference between a column density of 21cm radio observation and that from X-ray spectra analysis is considered as a cold medium around the M87 jet. A neutral density around the jet can be calculated with a size of an emission region. A neutral density $N$ (cm$^{-3}$) around each knot is calculated by dividing the difference of column densities with $2R$. Here, $R$ is a emission radius for each knot.

### 4.1.2 Plasma density

A normalization of an APEC model is given as $10^{-14}$ x $n_e n_i V / 4\pi D_A{}^2(1+z)^2$. Here, $n_e$ is an electron density in units of cm$^{-3}$ and $n_i$ is an ion density in units of cm$^{-3}$. $V$ is a volume of the extracted region in units of cm$^3$. $D_A$ is an angular diameter distance to M87 in units of cm. $z$ is a redshift. An electron density $n_e$ is assumed to be equal to an ion density $n_i$ and an ion density $n_i$ is derived as 11.0(+1.2, −1.4) cm$^{-3}$. When a metal abundance is fixed to 0.1 and 1 solar, a plasma density is 4.5(+0.2, −0.3)cm$^{-3}$ and 1.5($\pm$0.1)cm$^{-3}$, respectively. The error is 90% confidence level statistical error.

The ratio of a temperature is given by $T_2/T_1 = 2\gamma(\gamma-1)M^2/(\gamma+1)^2$ for strong shock $M$ >>1 (Longair 1992). Here, $T_1$ is a temperature outside the jet and $T_2$ is that in the jet. $M$ is a mach number of a shock wave. $\gamma$ is a ratio of a specific heat. When $\gamma$ is given as 5/3 for monatomic gas, $T_2/T_1 = 0.3 M^2$ >> 1. There is a possibility that the gas in the jet is heated by a shock and it is observed as thermal emission. The metal abundance of a thermal component is low compared with an intercluster medium of M87 which abundance is 1 solar (Belsole et al 2001). This suggests that the heating by a shock is occurred in the jet, not outside the jet. The ratio of pressure is given by $p_2/p_1 = 2\gamma M^2/(\gamma+1)$ for a strong shock $M$>>1(Longair 1992). Here, $p_1$ is a pressure outside the jet and $p_2$ is a pressure in the jet. When $\gamma$ is given as 5/3 for monatomic gas, $p_2/p_1 = 1.25 M^2$ >>1. There is no pressure valance between the jet and an interstellar medium. The ratio of a density is given by $\rho_2/\rho_1 = (\gamma+1)/(\gamma-1)$ for a strong shock $M$>>1 (Longair 1992). Here, $\rho_1$ is a material density outside the jet and $\rho_2$ is that in the jet. When $\gamma$ is given as 5/3 for monatomic gas, $\rho_2/\rho_1 = 4$. It is possible that a plasma density is comparable with a neutral density of a few cm$^{-3}$. The absorption in a cold medium and a soft excess are related. The observed normalization of a thermal component is a lower limit if there is a cold medium around the knot A.

The rotation measure for the knot A is RM 1000 rad $m^{-2}$ (Peng et al. 2024). RM is given as $8.12 \times 10^{3}\, n_e B_{//} L$ (Longair 1992). Here, $n_e$ is a plasma density in units of $m^{-3}$, $B_{//}$ is a magnetic field pallarel to the line of sight in units of T and $L$ is a size of a region in units of pc. When magnetic field is assumed to be 1 mG, a plasma density is given as $7.9 \times 10^{-3}$ $cm^{-3}$ with a size of 156 pc for the knot A. This value differs from X-ray energy spectra by an order of 3. If plasma density is correct, magnetic field $B_{//}$ is derived to be 1 $\mu$G for knot A

4.1.3 Density of an accelerated electron

Synchrotron emissivity J are given by $2.344 \times 10^{-25}$ x $a(p)\, B^{(P+1)/2}\, K'\, (3.217 \times 10^{17}/\ \nu)^{(p-1)/2}$ W $m^{-3}$ $Hz^{-1}$ (Longair 1992). Here, a density of an accelerated electron is given by $dN_e/dE = K'\, E^{-p}$ $m^{-3}$ $GeV^{-1}$. $E$ is in units of GeV and $K'$ is in units of $m^{-3}$ $GeV^{p-1}$. $B$ is a magnetic field in units of T. A magnetic field are reported from different analyses. An observation of a synchrotron self-absorption of the nucleus at 43 GHz sets a limit on a magnetic field from 1 G to 10 G (Kino et al., 2014). A timescale of an energy loss of synchrotron emission is proportional to $1/B^2$ (Longair 1992). Variability in X-ray is different for a different part of the jet as shown in figure 2. An X-ray energy loss in the knot HST-1 is consistent with an $E^2$ energy loss of synchrotron emission and a magnetic field of 0.6 mG for $\delta$=5 is obtained (Harris et al. 2009). For knot A, non flux variability over 18 years suggests magnetic field is weaker than that of HST-1. The magnetic field parallel to the line of sight is 1 $\mu$G for knot A, but total magnetic field is unknown. Study of magnetic field from synchrotron energy loss time scale is a future work for knot D and knot A. Therefore, calculation is done only for nucleus and HST-1. A magnetic field of 1 G is used for the nucleus and 1 mG is used for HST-1. An index $p$ is given by $2\alpha - 1$($\alpha$ is the photon index). A numerical parameter $a(p)$ is a constant that depends on an index $p$. a(3.0) is 0.269, a(3.5) is 0.217 and a(4) is 0.186 (Longair 1992). $\nu$ is a frequency in units of Hz. A 1 keV flux of synchrotron emission is given by J $V\, \delta\, \Gamma^2/\ 4\pi D^2$. Here, $V$ is a volume of the extracted region in units of $cm^3$ which is given as $4\pi R^3/3$. $R$ is a radius of the extracted region in units of cm as shown in table 1. Here, $D$ is the distance of M87. Here, $\delta$ is a beaming factor and $\Gamma$ is a Lorentz factor. There is a beaming effect for accelerated electrons. A time duration in an observer system is $1/\delta$ time as long as that in a jet system. A solid angle in an observer system is $1/\Gamma^2$ times as large as that in a jet system. An apparent velocity for the knot HST-1 is 6.6 $c$ (Thimmappa et al 2024) which means an inclination angle of $\theta = 19^{\circ}$ and a Lorentz factor of $\Gamma$=3 by $\delta = 1/\Gamma(1 - \beta\cos\theta)$. $K'$

is derived with $\delta=6$ and $\Gamma=3$ for all parts of the jet. $K'$ is small for the nucleus and large for HST-1. This is due to a difference in a magnetic field against almost same X-ray flux.

### 4.2 Non thermal bremsstrahlung

It is considered that there is an interaction between accelerated particles (proton, electron) and an interstellar medium. There may be the interaction between accelerated protons and an interstellar medium. Gamma ray from a pion decay can be detected from a shape of energy spectra which peak at 70 MeV. However, energy spectra below 100 MeV for M87 has not been published with *Fermi*. A high density of accelerated electrons from HST-1 is obtained. The flux of non thermal bremsstrahlung as an interaction between accelerated electrons and an interstellar medium is calculated.

#### 4.2.1 Calculated flux with *Chandra*

Flux of non thermal bremsstrahlung for a cold material I is given as $bN \times K' \epsilon^{-p} / (p-1)$ ph/m$^3$/s/GeV (Longair 1992). Here, b is $1.03 \times 10^{-21}$ m$^3$/s. $N$ is a neutral density in unit of m$^{-3}$. $\epsilon$ is an energy in units of GeV. A number density of accelerated electrons is given as $dN_e/dE = K'(E)^{-p}$ m$^{-3}$ GeV$^{-1}$. $K'$ is in units of m$^{-3}$ GeV$^{p-1}$. $E$ is in units of GeV. An index $p$ is given by $2\alpha-1$ ($\alpha$ is the photon index). A total energy loss rate $dE/dt$ of a non-thermal bremsstrahlung for fully ionized materials differs from that for cold materials by a factor of 4 for $\Gamma=3$ (Longair 1992). Therefore, the same formula is used for fully ionized materials. Flux is given by $I\ V \delta\ \Gamma^2 / 4\pi D^2$. Calculated flux of non thermal bremsstrahlung is independent of $\delta$ and $\Gamma$ because $\delta$ and $\Gamma$ are used when $K'$ is calculated from 1 keV flux of a power law. The calculated flux of non thermal bremsstrahlung from an X-ray result is shown in table 11.

There is a peak in the energy spectra of $E^2\ dN/dE$ vs. $E$ from a radio to an X-ray, where is a ultra violet. This break is not due to a synchrotron self absorption which is shown in a radio band. When this energy spectra is assumed as synchrotron emission, the derived spectral index of an electron distribution $dN_e/dE$ has a break at 460(+20, −30) GeV and an index of an electron distribution $dN_e/dE$ is 1.77(±0.07) in a low energy for HST-1(L state in Sun et al. 2018). The synchrotron emission from electrons which have an energy $E$ peaks at an energy of $(E/m_e c^2)^2 (eB/2\pi m_e)$ (Longair 1992). Here, $B$ is a magnetic field. 1 keV photon is due to 5 TeV electron with a magnetic field of 1 mG. Therefore, the calculated flux of non thermal bremsstrahlung from an X-ray result should be modified with the effect of a break of an index. The modified flux of non thermal bremsstrahlung is shown in table 11.

### 4.2.2 Comparison with the observed flux with *Fermi*

The total observed flux from 100 MeV to 100 GeV with *Fermi* is $(1.79\pm0.10)\times10^{-8}$ ph/s/cm$^2$(Cao et al. 2024). The total flux from 100 MeV to 100 GeV of non thermal bremsstrahlung from the knot HST-1 (all data) is calculated as $2.4(+2.6, -2.3)\times10^{-14}$ ph/s/cm$^2$. The non thermal bremsstrahlung flux from HST-1 is smaller than the observed flux with *Fermi* by an order of 6. The calculate flux of non thermal bremsstrahlung from nucleus is small even without modification. There is no contribution of non thermal bremsstrahlung from nucleus and HST-1 to the observed flux with *Fermi*.

The non thermal bremsstrahlung flux is largely dependent on a magnetic field. If magnetic field is less than 10 $\mu$G for knot A, I suggests that there is some contribution of non thermal bremsstrahlung flux from knot A to observed flux with *Fermi*. The study of magnetic field for knot A from synchrotron energy loss time scale is a future work.

## 5 Conclusion

I study both a thermal component and a non thermal component as an interaction between the jet and an interstellar medium of M87 by *Chandra*. It is confirmed that an X ray energy spectra for the nucleus and the knot D are well described with a power law model. It is found that a power law model is rejected statistically for HST-1 and knot A and an X ray energy spectra is well described with a combination model of a power law and a thermal bremsstrahlung of temperature of 0.2 keV and a metal abundance of 0.00. There is a possibility that gas in the jet is heated by a shock and it is observed as a thermal component. Flux of non thermal bremsstrahlung is calculated from an X ray result. It is found that there is no contribution of non thermal bremsstrahlung from nucleus and the knot HST-1 to the observed flux with *Fermi*.

Acknowledgement

I thank prof. Makishima K. and Dr. Matsushita K. for useful comments about data analysis. This research has made use of data obtained from the *Chandra Data Archive*, and software provided by the *Chandra X-ray Center* (*CXC*) in the application packages *CIAO*.

Table 1 The extracted region for each knot of the M87 jet for obsID 18232. 1" is 78 pc.

| | RA(J2000) | DEC(J2000) | Distance from nucleus | radius of Region |
|---|---|---|---|---|
| nucleus | $12^h30^m49^s.46$ | $12^o$ 23' 28".0 | | 0".5 |
| HST-1 | $12^h30^m49^s.38$ | $12^o$ 23' 28".4 | 1".2 (94 pc) | 0".6 |
| D | $12^h30^m49^s.27$ | $12^o$ 23' 28".9 | 2".9 (226 pc) | 0".75 |
| A | $12^h30^m48^s.65$ | $12^o$ 23' 32".3 | 12".8 (998 pc) | 1".0 |

Table 2.a The observation log used for the nucleus with 5% pile up fraction

| obsID | PI | obs date | number |
|---|---|---|---|
| former data set | | | |
| 16042 | | 2013.12 | 1 |
| 17056~17057 | | 2014.12~2015.3 | 2 |
| exposure time | 14ks | | |
| latter data set | | | |
| 18809~18813 | Cheng | 2016.3 | 5 |
| 18232~18233 | Russell | 2016.2~2016.4 | 2 |
| 18781~18783 | | 2016.2~2016.4 | 3 |
| 18836~18838 | | 2016.4 | 3 |
| 18856 | | 2016.6 | 1 |
| 20034 | Neilsen | 2017.4 | 1 |
| 19457~19458 | Wong | 2017.2 | 2 |
| exposure time | 345ks | | |
| total exposure time | 358ks | | |

Table 2.b The observation log used for nucleus with 10% pile up fraction

| obsID | PI | obs date | number |
|---|---|---|---|
| former data set | | | |
| 3084 | Harris | 2002.2 | 1 |
| 3976 | Harris | 2002.12 | 1 |
| 6303 | Biretta | 2006.5 | 1 |
| 8581 | Biretta | 2008.8 | 1 |
| 10282~10287 | Harris | 2008.11~2009.6 | 6 |
| 11513~11520 | Harris | 2010.4~2010.5 | 8 |
| 13964 | Harris | 2011.12 | 1 |
| 14973~14974 | | 2012.12~2013.3 | 2 |
| 16042 | | 2013.12 | 1 |
| 17056~17057 | | 2014.12~2015.3 | 2 |
| exposure time | 111ks | | |
| latter data set | | | |
| 18809~18813 | Cheng | 2016.3 | 5 |
| 18232~18233 | Russell | 2016.2~2016.4 | 2 |
| 18781~18783 | | 2016.2~2016.4 | 3 |
| 18836~18838 | | 2016.4 | 3 |
| 18856 | | 2016.6 | 1 |
| 20034 | Neilsen | 2017.4 | 1 |
| 19457~19458 | Wong | 2017.2 | 2 |
| exposure time | 345ks | | |
| total exposure time | 456ks | | |

Table 2.c The observation log used for the knot HST-1 with 5% pile up fraction

| obsID | PI | obs date | number |
|---|---|---|---|
| former data set | | | |
| 11518 | Harris | 2010.5 | 1 |
| 13964 | Harris | 2011.12 | 1 |
| 14973~14974 | | 2012.12~2013.3 | 2 |
| 16042 | | 2013.12 | 1 |
| 17056~17057 | | 2014.12~2015.3 | 2 |
| exposure time | 32ks | | |
| latter data set | | | |
| 18809~18813 | Cheng | 2016.3 | 5 |
| 18232~18233 | Russell | 2016.2~2016.4 | 2 |
| 18781~18783 | | 2016.2~2016.4 | 3 |
| 18836~18838 | | 2016.4 | 3 |
| 18856 | | 2016.6 | 1 |
| 20034 | Neilsen | 2017.4 | 1 |
| 19457~19458 | Wong | 2017.2 | 2 |
| exposure time | 345ks | | |
| total exposure time | 377ks | | |

Table 2.d The observation log used for HST-1 with 10% pile up fraction

| obsID | PI | obs date | number |
|---|---|---|---|
| former data set | | | |
| 3084 3086 | Harris | 2002.2~2002.3 | 2 |
| 10282~10288 | Harris | 2008.11~2009.12 | 7 |
| 11512~11520 | Harris | 2010.4~2010.5 | 9 |
| 13964 | Harris | 2011.12 | 1 |
| 14973~14974 | | 2012.12~2013.3 | 2 |
| 16042 | | 2013.12 | 1 |
| 17056~17057 | | 2014.12~2015.3 | 2 |
| exposure time | 111ks | | |
| latter data set | | | |
| 18809~18813 | Cheng | 2016.3 | 5 |
| 18232~18233 | Russell | 2016.2~2016.4 | 2 |
| 18781~18783 | | 2016.2~2016.4 | 3 |
| 18836~18838 | | 2016.4 | 3 |
| 18856 | | 2016.6 | 1 |
| 20034 | Neilsen | 2017.4 | 1 |
| 19457~19458 | Wong | 2017.2 | 2 |
| exposure time | 345ks | | |
| total exposure time | 456ks | | |

Table 2.e The observation log used for the knot D

| obsID | PI | obs date | number |
|---|---|---|---|
| former data set | | | |
| 1808 | Wilson | 2000.7 | 1 |
| 3084~3088 | Harris | 2002.2~2002.7 | 5 |
| 3975~3982 | Harris | 2002.11~2003.8 | 8 |
| 4917~4919 | Birreta | 2003.11~2004.8 | 3 |
| 6301~6303,6305 | Birreta | 2006.2~2006.8 | 4 |
| 7352 | Birreta | 2007.5 | 1 |
| 8510~8517 | Harris | 2007.2~2007.3 | 8 |
| 8575~8581 | Birreta | 2007.11~2008.8 | 7 |
| 10282~10288 | Harris | 2008.11~2009.12 | 7 |
| 11512~11520 | Harris | 2010.4~2010.5 | 9 |
| 13964~13965 | Harris | 2011.12~2012.2 | 2 |
| 14973~14974 | | 2012.12~2013.3 | 2 |
| 16042~16043 | | 2013.12~2014.4 | 2 |
| 17056~17057 | | 2014.12~2015.3 | 2 |
| exposure time | 294ks | | |
| latter data set | | | |
| 18809~18813 | Cheng | 2016.3 | 5 |
| 18232~18233 | Russell | 2016.2~2016.4 | 2 |
| 18781~18783 | | 2016.2~2016.4 | 3 |
| 18836~18838 | | 2016.4 | 3 |
| 18856 | | 2016.6 | 1 |
| 20034~20035 | Neilsen | 2017.4 | 2 |
| 19457~19458 | Wong | 2017.2 | 2 |
| 21075~21076 | | 2018.4 | 2 |
| 20488~20489 | Cheng | 2018.1~2018.3 | 2 |
| exposure time | 385ks | | |
| total exposure time | 679ks | | |

Table 2.f The observation log used for knot A

| obsID | PI | obs date | number |
| --- | --- | --- | --- |
| former data set | | | |
| 1808 | Wilson | 2000.7 | 1 |
| 3084~3088 | Harris | 2002.2~2002.7 | 5 |
| 3975~3982 | Harris | 2002.11~2003.8 | 8 |
| 4917~4919 | Birreta | 2003.11~2004.2 | 3 |
| 5740 | Birreta | 2005.4 | 1 |
| 6301~6303,6305 | Birreta | 2006.2~2006.8 | 4 |
| 7351 7352 | Birreta | 2007.3~2007.5 | 2 |
| 8510~8517 | Harris | 2007.2~2007.3 | 8 |
| 8575~8581 | Birreta | 2007.11~2008.8 | 7 |
| 10282~10288 | Harris | 2008.11~2009.12 | 7 |
| 11512~11520 | Harris | 2010.4~2010.5 | 9 |
| 13964~13965 | Harris | 2011.12~2012.2 | 2 |
| 14973~14974 | | 2012.12~2013.3 | 2 |
| 16042~16043 | | 2013.12~2014.4 | 2 |
| 17056~17057 | | 2014.12~2015.3 | 2 |
| exposure time | 304ks | | |
| latter data set | | | |
| 18809~18813 | Cheng | 2016.3 | 5 |
| 18232~18233 | Russell | 2016.2~2016.4 | 2 |
| 18781~18783 | | 2016.2~2016.4 | 3 |
| 18836~18838 | | 2016.4 | 3 |
| 18856 | | 2016.6 | 1 |
| 20034~20035 | Neilsen | 2017.4 | 2 |
| 19457~19458 | Wong | 2017.2 | 2 |
| 21075~21076 | | 2018.4 | 2 |
| 20488~20489 | Cheng | 2018.1~2018.3 | 2 |
| exposure time | 385ks | | |
| total exposure time | 689ks | | |

Table 3 The fitting results with a power law model for a former data set of each part of a jet. A photo absorption is applied to power law model and is set free. An error is a 90% confidence level statistical error. ( )* is pile up fraction.

| | nucleus(<5%)* | nucleus(<10%)* | HST-1(<5%)* |
|---|---|---|---|
| $N_{H}$(x$10^{20}$ cm$^{-2}$) | $6.58^{+4.32}_{-4.04}$ | $6.32^{+1.06}_{-1.04}$ | $1.66^{+3.42}_{-1.66}$ |
| photon index | $2.00^{+0.14}_{-0.13}$ | $2.12^{+0.03}_{-0.04}$ | $2.47^{+0.15}_{-0.11}$ |
| 1keV flux(ph/cm$^2$/s/keV) | $4.26^{+0.57}_{-0.50}$ (x$10^{-4}$) | $5.27^{+0.18}_{-0.18}$ (x$10^{-4}$) | $2.26^{+0.27}_{-0.18}$(x$10^{-4}$) |
| $\chi^2$/d.o.f (d.of) | 0.876(103) | 1.035(287) | 1.080(109) |
| | HST-1(<10%)* | D | A |
| $N_{H}$(x$10^{20}$ cm$^{-2}$) | $1.93^{+1.17}_{-1.15}$ | $0.24^{+1.17}_{-0.24}$ | 0.00 |
| photon index | $2.43^{+0.05}_{-0.05}$ | $2.23^{+0.05}_{-0.03}$ | $2.43^{+0.02}_{-0.01}$ |
| 1keV flux(ph/cm$^2$/s/keV) | $4.18^{+0.17}_{-0.16}$(x$10^{-4}$) | $1.11^{+0.05}_{-0.02}$(x$10^{-4}$) | $2.20^{+0.02}_{-0.02}$(x$10^{-4}$) |
| $\chi^2$ /d.o.f (d.o.f) | 0.949(223) | 1.004(238) | 1.357(294) |

Table 4 The fitting result with both power law model and combination model of a power law and an APEC for a former data set of the knot A. A photo absorption is applied to power law model and APEC model, and is fixed to a 21 cm radio observation value. Here, PL is a power law. An error is a 90% confidence level statistical error.

| Model | | A |
|---|---|---|
| PL | photon index | $2.49^{+0.02}_{-0.02}$ |
| | 1keV flux(ph/cm$^2$/s/keV) | $2.29^{+0.02}_{-0.02}$ (x$10^{-4}$) |
| | $\chi^2$/d.o.f (d.of) | 1.567(295) |
| PL+APEC | photon index | $2.28^{+0.04}_{-0.03}$ |
| | 1 keV flux(ph/cm$^2$/s/keV) | $2.00^{+0.06}_{-0.06}$(x$10^{-4}$) |
| | $kT$(keV) | $0.19^{+0.03}_{-0.03}$ |
| | abundance | $0.00^{+0.01}$ |
| | normalization | $2.65^{+1.19}_{-0.82}$(x$10^{-3}$) |
| | $\chi^2$/d.o.f(d.o.f) | 0.953(292) |

Table 5 The fitting result with power law model for a latter data set of each part of a jet. A photo absorption is applied to power law model and is set free. An error is a 90% confidence level statistical error.

| | nucleus | HST-1 |
|---|---|---|
| $N_H$(x$10^{20}$ cm$^{-2}$) | $8.89^{+1.22}_{-1.21}$ | $1.09^{+1.76}_{-1.09}$ |
| photon index | $2.19^{+0.04}_{-0.03}$ | $2.30^{+0.06}_{-0.05}$ |
| 1keV flux(ph/cm$^2$/s/keV) | $2.89^{+0.10}_{-0.09}$ (x$10^{-4}$) | $1.18^{+0.07}_{-0.05}$ (x$10^{-4}$) |
| $\chi^2$/d.o.f (d.of) | 1.063(320) | 1.087(228) |
| | D | A |
| $N_H$(x$10^{20}$ cm$^{-2}$) | $1.87^{+1.39}_{-1.37}$ | 0.00 |
| photon index | $2.32^{+0.04}_{-0.04}$ | $2.38^{+0.02}_{-0.02}$ |
| 1keV flux(ph/cm$^2$/s/keV) | $1.55^{+0.07}_{-0.06}$(x$10^{-4}$) | $1.95^{+0.03}_{-0.02}$(x$10^{-4}$) |
| $\chi^2$ /d.o.f (d.o.f) | 0.981(272) | 1.098(298) |

Table 6 The fitting result with a power law model for latter data set for knot A. A photo absorption is applied to a power law model and is fixed to a 21 cm radio observation value. PL is a power law model. An error is a 90% confidence level statistical error.

| Model | | A |
|---|---|---|
| PL | photon index | $2.42^{+0.02}_{-0.02}$ |
| | 1keV flux(ph/cm$^2$/s/keV) | $2.02^{+0.03}_{-0.02}$ (x$10^{-4}$) |
| | $\chi^2$/d.o.f (d.of) | 1.179(299) |

Table 7 The fitting result with a power law model for all data of each part of a jet. A photo absorption is applied to a power law model and is set free. An error is a 90% confidence level statistical error. ( )* is pile up fraction.

| | nucleus(<5%)* | nucleus(<10%)* | HST-1(<5%)* |
|---|---|---|---|
| $N_H$(x$10^{20}$ cm$^{-2}$) | $8.48^{+1.17}_{-1.15}$ | $4.10^{+0.77}_{-0.76}$ | 0.00 |
| photon index | $2.18^{+0.04}_{-0.03}$ | $2.15^{+0.02}_{-0.03}$ | $2.33^{+0.02}_{-0.03}$ |
| 1keV flux(ph/cm$^2$/s/keV) | $2.95^{+0.10}_{-0.09}$ (x$10^{-4}$) | $3.42^{+0.08}_{-0.08}$ (x$10^{-4}$) | $1.28^{+0.03}_{-0.02}$(x$10^{-4}$) |
| $\chi^2$/d.o.f (d.of) | 1.069(326) | 1.147(364) | 0.988(241) |
| | HST-1(<10%)* | D | A |
| $N_H$(x$10^{20}$ cm$^{-2}$) | $0.00^{+0.10}$ | $2.62^{+0.86}_{-0.84}$ | 0.00 |
| photon index | $2.51^{+0.02}_{-0.02}$ | $2.28^{+0.03}_{-0.03}$ | $2.43^{+0.01}_{-0.02}$ |
| 1keV flux(ph/cm$^2$/s/keV) | $2.17^{+0.03}_{-0.02}$(x$10^{-4}$) | $1.36^{+0.04}_{-0.04}$(x$10^{-4}$) | $2.12^{+0.01}_{-0.02}$(x$10^{-4}$) |
| $\chi^2$ /d.o.f (d.o.f) | 1.191(287) | 1.090(315) | 1.485(354) |

Table 8 The fitting result with a power law model and combination model of power law and APEC for all data of HST-1 and the knot A. A photo absorption is applied to a power law model and an APEC model, and fixed to a 21 cm radio value. PL is a power law model. An error is a 90% confidence level statistical error.

| Model | | HST-1 | HST-1(10%) | A |
|---|---|---|---|---|
| PL | photon index | $2.36^{+0.03}_{-0.03}$ | $2.55^{+0.02}_{-0.02}$ | $2.47^{+0.02}_{-0.01}$ |
| | 1keV flux(ph/cm$^2$/s/keV) | $1.33^{+0.02}_{-0.02}$ (x$10^{-4}$) | $2.26^{+0.02}_{-0.02}$ (x$10^{-4}$) | $2.20^{+0.02}_{-0.01}$ (x$10^{-4}$) |
| | $\chi^2$/d.o.f (d.of) | 1.000(242) | 1.315(288) | 1.771(355) |
| PL+APEC | photon index | | $2.40^{+0.03}_{-0.04}$ | $2.28^{+0.03}_{-0.03}$ |
| | 1 keV flux(ph/cm$^2$/s/keV) | | $2.02^{+0.06}_{-0.07}$(x$10^{-4}$) | $1.91^{+0.04}_{-0.03}$(x$10^{-4}$) |
| | $kT$(keV) | | $0.22^{+0.03}_{-0.01}$ | $0.20^{+0.03}_{-0.02}$ |
| | abundance | | 0.06 | 0.00 |
| | normalization | | $5.22^{+6.58}_{-5.14}$(x$10^{-4}$) | $2.32^{+0.55}_{-0.60}$(x$10^{-3}$) |
| | $\chi^2$/d.o.f(d.o.f) | | 0.941(285) | 0.913(352) |

Table 9 The fitting result with a combination model of a power law and APEC for all data of knot A. photo absorption is applied to a power law model and APEC model, and is fixed to 21 cm observation value. Metal abundance is fixed to 0.1 solar and 1 solar. An error is a 90% confidence level statistical error.

| | | |
|---|---|---|
| abundance 0.1 solar | photon index | $2.32^{+0.02}_{-0.02}$ |
| | 1 keV flux(ph/cm$^2$/s/keV) | $1.99^{+0.03}_{-0.03}$(x$10^{-4}$) |
| | $kT$(keV) | $0.20^{+0.01}_{-0.01}$ |
| | normalization | $3.77^{+0.38}_{-0.39}$(x$10^{-4}$) |
| | $\chi^2$/d.o.f(d.o.f) | 1.001(353) |
| abundance 1 solar | photon index | $2.33^{+0.02}_{-0.02}$ |
| | 1 keV flux(ph/cm$^2$/s/keV) | $2.00^{+0.03}_{-0.02}$(x$10^{-4}$) |
| | $kT$(keV) | $0.20^{+0.01}_{-0.01}$ |
| | normalization | $4.33^{+0.44}_{-0.43}$(x$10^{-5}$) |
| | $\chi^2$/d.o.f(d.o.f) | 1.032(353) |

Table 10 The three physical parameters from X-ray fitting result for each data sets. PL is a power law model. $(\ )^{*}$ is pile up fraction. An error is a 90% confidence level statistical error.

| | Best Model | Neutral density $N$ ($cm^{-3}$) | plasma density $n_i$ ($cm^{-3}$) | $p=2\alpha-1$ | $K'$($m^{-3}GeV^{p-1}$) |
|---|---|---|---|---|---|
| former data set | | | | | |
| nucleus(<5%)$^{*}$ | PL | $2.2^{+1.8}_{-1.7}$ | | $3.00^{+0.28}_{-0.26}$ | $1.6^{+6.2}_{-1.3}$(x$10^{-7}$) |
| nucleus(<10%)$^{*}$ | PL | $2.1^{+0.4}_{-0.4}$ | | $3.24^{+0.06}_{-0.08}$ | $5.8^{+3.9}_{-1.9}$ (x$10^{-7}$) |
| HST-1(<5%)$^{*}$ | PL | $0.1^{+1.2}_{-0.1}$ | | $3.94^{+0.30}_{-0.22}$ | $1.1^{+11.9}_{-0.9}$ (x$10^{2}$) |
| HST-1(<10%)$^{*}$ | PL | $0.2^{+0.4}_{-0.2}$ | | $3.86^{+0.10}_{-0.10}$ | $1.1^{+1.4}_{-0.6}$ (x$10^{2}$) |
| D | PL | 0.0 | | $3.46^{+0.10}_{-0.06}$ | |
| A | PL+APEC | >0.0 | $11.8^{+2.7}_{-1.9}$ | $3.56^{+0.08}_{-0.06}$ | |
| latter data set | | | | | |
| nucleus | PL | $3.2^{+0.5}_{-0.5}$ | | $3.38^{+0.08}_{-0.06}$ | $7.3^{+3.7}_{-1.9}$(x$10^{-7}$) |
| HST-1 | PL | $0.0^{+0.5}$ | | $3.60^{+0.12}_{-0.10}$ | $3.3^{+5.6}_{-1.9}$ |
| D | PL | $0.2^{+0.4}_{-0.2}$ | | $3.64^{+0.08}_{-0.08}$ | |
| A | PL | 0.0 | | $3.84^{+0.04}_{-0.04}$ | |
| all data | | | | | |
| nucleus(<5%)$^{*}$ | | $3.0^{+0.5}_{-0.5}$ | | $3.36^{+0.08}_{-0.06}$ | $6.9^{+3.1}_{-1.8}$(x$10^{-7}$) |
| nucleus(<10%)$^{*}$ | PL | $1.2^{+0.3}_{-0.3}$ | | $3.30^{+0.04}_{-0.06}$ | $6.1^{+1.3}_{-2.4}$(x$10^{-7}$) |
| HST-1(<5%)$^{*}$ | PL | 0.0 | | $3.72^{+0.06}_{-0.06}$ | $9.5^{+8.6}_{-3.7}$ |
| HST-1(<10%)$^{*}$ | PL+APEC | >0.0 | $12.6^{+3.5}_{-11.3}$ | $3.80^{+0.06}_{-0.08}$ | $3.2^{+2.1}_{-1.8}$(x10) |
| D | PL | $0.4^{+0.2}_{-0.3}$ | | $3.56^{+0.06}_{-0.06}$ | |
| A | PL+APEC | >0.0 | $11.0^{+1.2}_{-1.4}$ | $3.56^{+0.06}_{-0.06}$ | |

Table 11 The flux of non thermal bremsstrahlung at 1 GeV for nucleus and the knot HST-1. The error is 90% confidence level statistical error. $(\ )^{*}$ is pile up fraction.

| | unmodified flux | modified flux with a break |
|---|---|---|
| | (ph $cm^{-2}$ $s^{-1}$) | (ph $cm^{-2}$ $s^{-1}$) |
| | (lower limit, upper limit) | (lower limit, upper limit) |
| former data set | | |
| nuclues(<5%)$^{*}$ | 3.8x$10^{-19}$(2.2x$10^{-20}$,3.0x$10^{-18}$) | |
| nucleus(<10%)$^{*}$ | 1.2x$10^{-18}$(6.7x$10^{-19}$, 2.3x$10^{-18}$) | |
| HST-1(<5%)$^{*}$ | 2.0x$10^{-11}$(0.0, 2.2x$10^{-9}$) | 3.3x$10^{-17}$(0.0, 1.1x$10^{-15}$) |
| HST-1(<10%)$^{*}$ | 3.4x$10^{-11}$(0.0, 2.1x$10^{-10}$) | 9.3x$10^{-17}$(0, 5.5x$10^{-16}$) |
| latter data set | | |
| nucleus | 2.1x$10^{-18}$(1.4x$10^{-18}$, 3.5x$10^{-18}$) | |
| HST-1 | 0.0(0.0, 6.9x$10^{-12}$) | 0.0(0.0, 7.7x$10^{-17}$) |
| all data | | |
| nucleus(<5%)$^{*}$ | 1.9x$10^{-18}$(1.2x$10^{-18}$, 3.1x$10^{-18}$) | |
| nucleus(<10%)$^{*}$ | 6.9x$10^{-19}$(3.0x$10^{-19}$, 1.0x$10^{-18}$) | |
| HST-1(<5%)$^{*}$ | 0.0 | 0.0 |
| HST-1(<10%)$^{*}$ | 6.0x$10^{-10}$(2.8x$10^{-11}$, 1.2x$10^{-9}$) | 2.4x$10^{-15}$(1.1x$10^{-16}$, 5.7x$10^{-15}$) |

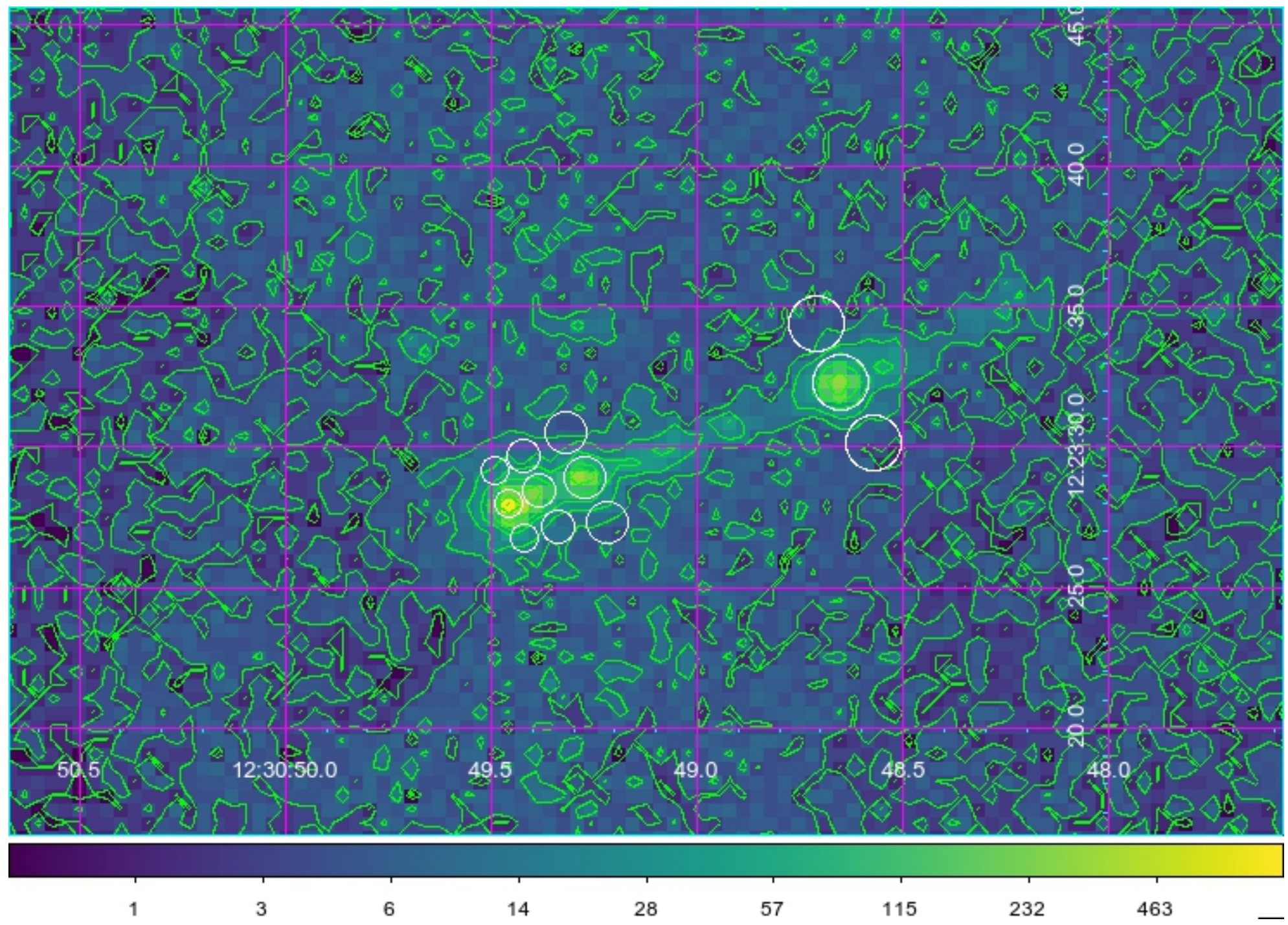

Figure 1 An X-ray image of the M87 jet for obsID 18232 with *Chandra*. From left, the nucleus, the knot HST-1, the knot D, the knot A, and each background region, respectively.

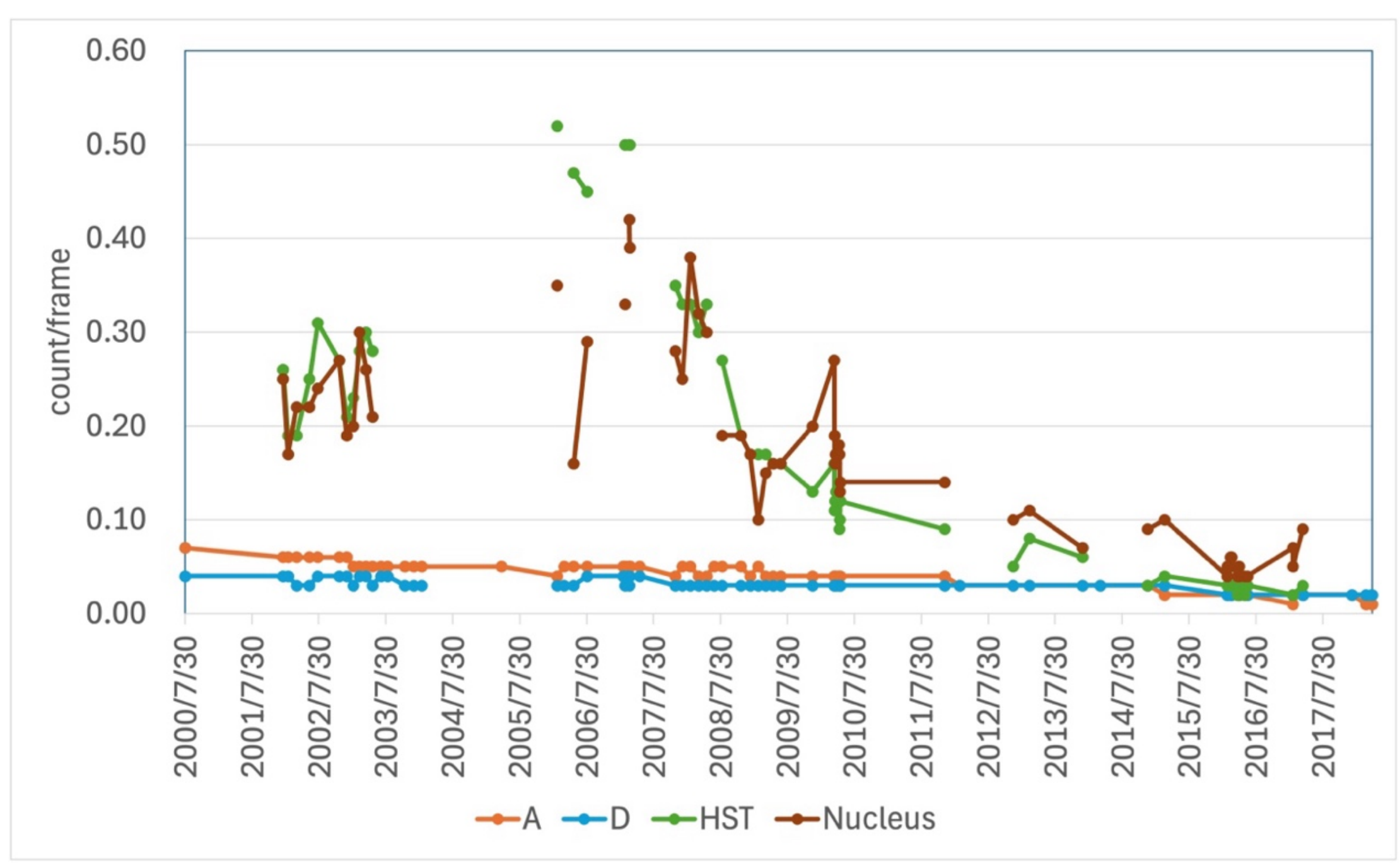

Figure 2 The count / frame from 2000 to 2018 for each part of jet. The count / frame = 0.1, 0.2 corresponds to 5%, 10% pile up fraction, respectively.

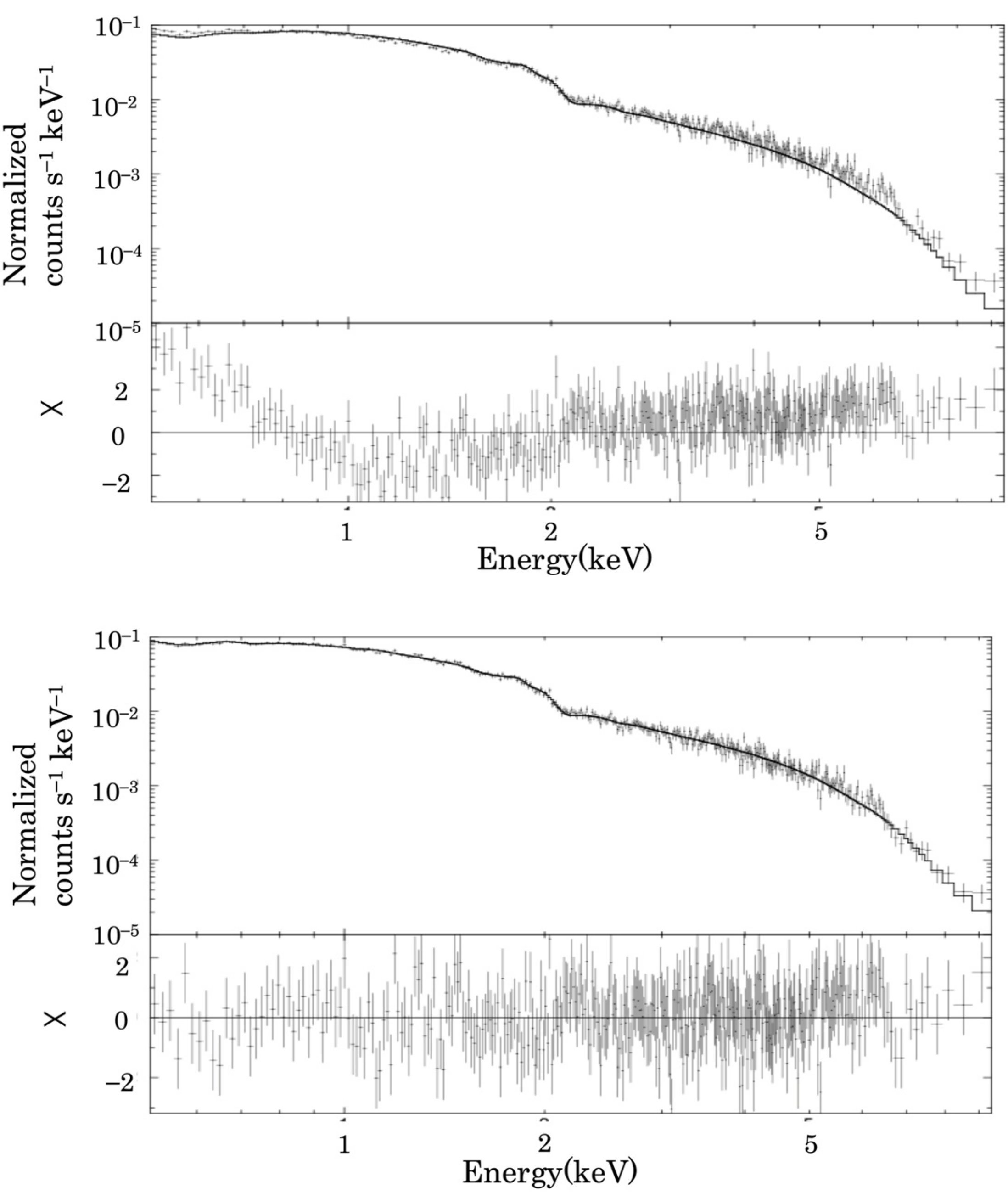

Figure 3 The fitted X-ray energy spectra with a power law model (top) and a combination of a power law and an APEC (bottom) for all data of the knot A. A photo absorption is applied to a power law model and APEC model, and fixed to a 21 cm radio observation value.